\documentclass[prb,amsmath,amssymb,twocolumn,showpacs,floatfix,superscriptaddress]{revtex4}
\usepackage{graphicx}
\usepackage{dcolumn}
\usepackage{bm}

\begin{document}

\title{Zero-Field $\mu$SR Search for a Time-Reversal-Symmetry-Breaking Mixed Pairing State
in Superconducting Ba$_{1-x}$K$_x$Fe$_2$As$_2$}

\author{Z.~Lotfi Mahyari}
\affiliation{Department of Physics, Simon Fraser University, Burnaby, British Columbia V5A 1S6, Canada}
\author{A.~Cannell}
\affiliation{Department of Physics, Simon Fraser University, Burnaby, British Columbia V5A 1S6, Canada}
\author{C.~Gomez}
\affiliation{Department of Physics, Simon Fraser University, Burnaby, British Columbia V5A 1S6, Canada}
\author{S.~Tezok}
\affiliation{Department of Physics, Simon Fraser University, Burnaby, British Columbia V5A 1S6, Canada}
\author{A.~Zelati}
\altaffiliation[Permanent Address: ]{Department of Physics, University of Birjand, Iran}
\affiliation{Department of Physics, Simon Fraser University, Burnaby, British Columbia V5A 1S6, Canada}
\author{E.V.L. de Mello}
\affiliation{Instituto de F\'{i}sica, Universidade Federal Fluminense, Niter\'{o}i, RJ 24210-340, Brazil}
\author{J.-Q Yan} 
\affiliation{Materials Science and Technology Division, Oak Ridge National Laboratory, Oak Ridge, Tennessee 37831, USA}
\affiliation{Department of Materials Science and Engineering, University of Tennessee, Knoxville, Tennessee 37996, USA}
\author{D.G.~Mandrus}
\affiliation{Materials Science and Technology Division, Oak Ridge National Laboratory, Oak Ridge, Tennessee 37831, USA}
\affiliation{Department of Materials Science and Engineering, University of Tennessee, Knoxville, Tennessee 37996, USA}
\author{J.E.~Sonier}
\affiliation{Department of Physics, Simon Fraser University, Burnaby, British Columbia V5A 1S6, Canada}
\affiliation{Canadian Institute for Advanced Research, Toronto, Canada}

\date{\today}

\begin{abstract}
We report the results of a zero-field muon spin relaxation (ZF-$\mu$SR) study of superconducting 
Ba$_{1-x}$K$_x$Fe$_2$As$_2$ ($0.5 \! \leq \! x \! \leq \! 0.9$) in search of weak spontaneous internal magnetic fields 
associated with proposed time-reversal-symmetry breaking mixed pairing states.
The measurements were performed on polycrystalline samples, which do not exhibit the mesoscopic phase separation
previously observed in single crystals of Ba$_{1-x}$K$_x$Fe$_2$As$_2$. 
No evidence of spontaneous internal magnetic fields is found in any of the samples at temperatures down to $T \! \sim \! 0.02$~K.
\end{abstract}

\maketitle
 
The microscopic mechanism responsible for superconductivity in iron-based superconductors
manifests itself in the Cooper-pair wave function symmetry, and consequently the symmetry of the
superconducting ground state has been a central issue of investigation.\cite{Hirschfeld:11} In the 122 iron-based 
superconductor Ba$_{1-x}$K$_x$Fe$_2$As$_2$ there is evidence for a transformation of the pairing symmetry 
with hole doping. Near optimal hole-doping ($x \! \sim \! 0.4$) the pairing state is widely believed to 
be of $s_{\pm}$ symmetry, with a full superconducting (SC) gap occurring on hole Fermi surface (FS) 
pockets at the Brillouin zone (BZ) center ($\Gamma$ point), and a full SC gap of opposite sign present 
on electron FS pockets centered about the M($\pi$,0)/(0,$\pi$) point.\cite{Ding:11} Such an  
$s_{\pm}$-wave pairing state may be mediated by spin fluctuations.\cite{Mazin:08}

Currently being debated is the situation at strong doping, where the electron FS pockets essentially vanish.\cite{Sato:09}
Laser angle-resolved photoemission spectroscopy (ARPES) measurements indicate that KFe$_2$As$_2$ ($x \! = \! 1$) 
has a complicated SC gap structure, with full and nodal gaps on the inner and middle BZ-centered hole FS pockets, 
respectively.\cite{Okazaki:12}     
Meanwhile, magnetic penetration depth,\cite{Hashimoto:10} thermal 
transport,\cite{Dong:10,Reid:12,Wang:12} and specific heat\cite{Abdel:13} measurements on KFe$_2$As$_2$ 
favor a state of $d$-wave pairing symmetry. The latter results support calculations predicting
the evolution of Ba$_{1-x}$K$_{x}$Fe$_2$As$_2$ into a nodal $d$-wave superconductor
concomitant with the disappearance of the electron FS pockets.\cite{Thomale:11,Maiti:11}
Yet recent ARPES measurements of Ba$_{0.1}$K$_{0.9}$Fe$_2$As$_2$ show 
that despite such a drastic change in the FS topology, isotropic SC gaps consistent with $s$-wave symmetry 
persist on unaltered hole FS pockets at the $\Gamma$ point.\cite{Xu:13}
 
While it remains unclear whether or how Ba$_{1-x}$K$_x$Fe$_2$As$_2$ transforms from a nodeless $s$-wave to nodal 
$d$-wave superconductor at full doping, a change from one pure pairing symmetry state to another may
occur via an intermediate phase of mixed symmetry, where the pure states are nearly degenerate. 
In particular, a time-reversal symmetry breaking (TRSB) 
$s \! + \! id$ state has been predicted to occur over some unspecified range of $x$ between optimal 
and full hole doping.\cite{Lee:09} The possibility of a mixed $s \! + \! id$ symmetry state
in Ba$_{1-x}$K$_{x}$Fe$_2$As$_2$ and other iron-based superconductors has been considered in several 
subsequent theoretical works.\cite{Stanev:10,Platt:12,Fernandes:13} Recently it has been proposed
that pure KFe$_2$As$_2$ actually has $s_{\pm}$ pairing symmetry, but differs from Ba$_{1-x}$K$_x$Fe$_2$As$_2$
at optimal doping in that the $s$-wave gaps of opposite sign occur on the hole FS pockets at the $\Gamma$ 
point.\cite{Maiti:13} In this case there is the possibility of an intermediate $s \! + \! is$ state
between optimal and full doping where the $s$-wave gaps on the hole FS pockets transform from having the same
to opposite signs. Like the proposed $s \! + \! id$ state, the $s \! + \! is$ state
breaks time-reversal symmetry.        
 
\begin{figure}
\centering
\includegraphics[width=8.0cm]{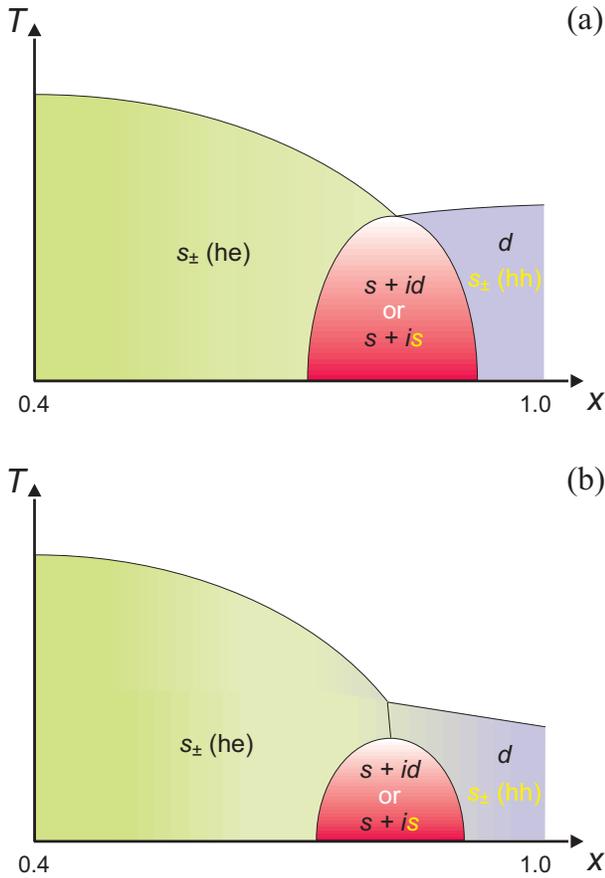}
\caption{(Color online) Schematic phase diagram of Ba$_{1-x}$K$_{x}$Fe$_2$As$_2$ as a function of temperature $T$ 
and $x$, showing a predicted mixed TRSB pairing state ($s \! + \! i d$ or $s \! + \! i s$) occurring somewhere 
between $x \! = \! 0.4$ and $x \! = \! 1$. The $s_{\pm}$ (he) state corresponds to
$s$-wave gaps of opposite sign on the hole (h) and electron (e) FS pockets. The $s_{\pm}$ (hh) state 
corresponds to $s$-wave gaps of opposite sign on the hole FS pockets at the $\Gamma$ point in the BZ
(in the absence of electron FS pockets) as proposed in Ref.~\onlinecite{Maiti:13}. In (a) the TRSB mixed symmetry 
state onsets at $T_c$ for a certain value of $x$, whereas in (b) it occurs significantly below $T_c$ 
irrespective of $x$. In (b) the pure $s_{\pm}$ (he) 
state evolves continuously into the pure $s_{\pm}$ (hh) state at temperatures above the 
$s \! + \! is$ state. On the other hand, a first-order transition (represented by a nearly vertical line)
occurs between the pure $s_{\pm}$ (he) and $d$-wave states at temperatures between $T_c$ and the onset of
the $s \! + \! id$ state.} 
\label{fig1}
\end{figure}

In a bulk superconductor with a TRSB order parameter, weak spontaneous currents are generated around 
impurities and lattice defects.\cite{Sigrist:98} This should occur even for the above mentioned TRSB multiband $s \! + \! is$ state.\cite{Sigrist} In such systems ZF-$\mu$SR has been demonstrated to
be an ideal local probe of the weak internal magnetic fields ($\sim \! 0.05$ to 1~G) produced by the spontaneous currents.
To date, weak internal fields compatible with a TRSB pairing state have been detected by ZF-$\mu$SR
in the superconducting phases of U$_{1-x}$Th$_x$Be$_{13}$,\cite{Heffner:90} Sr$_2$RuO$_4$,\cite{Luke:98}
PrOs$_4$Sb$_{12}$,\cite{Aoki:03} LaNiC$_2$,\cite{Hillier:09} PrPt$_4$Ge$_{12}$,\cite{Maisuradze:10}
Pr(Os$_{1-x}$Ru$_x$)$_4$Sb$_{12}$ and Pr$_{1-y}$La$_y$Os$_4$Sb$_{12}$,\cite{Shu:11}
LaNiGa$_2$,\cite{Hillier:12} and most recently in SrPtAs.\cite{Biswas:12}
The TRSB states that have been proposed for Ba$_{1-x}$K$_{x}$Fe$_2$As$_2$
do not necessarily occur immediately below the superconducting transition temperature ($T_c$), 
but may emerge at lower $T$.\cite{Lee:09,Maiti:13}
These two possibilities are depicted in Fig.~\ref{fig1}. Here we report a ZF-$\mu$SR search for a TRSB pairing 
state in Ba$_{1-x}$K$_{x}$Fe$_2$As$_2$ ($0.5 \! \leq \! x \! \leq \! 0.9$) at temperatures extending down 
to $T \! \sim \! 0.02$~K.

\begin{figure}
\centering
\includegraphics[width=8.0cm]{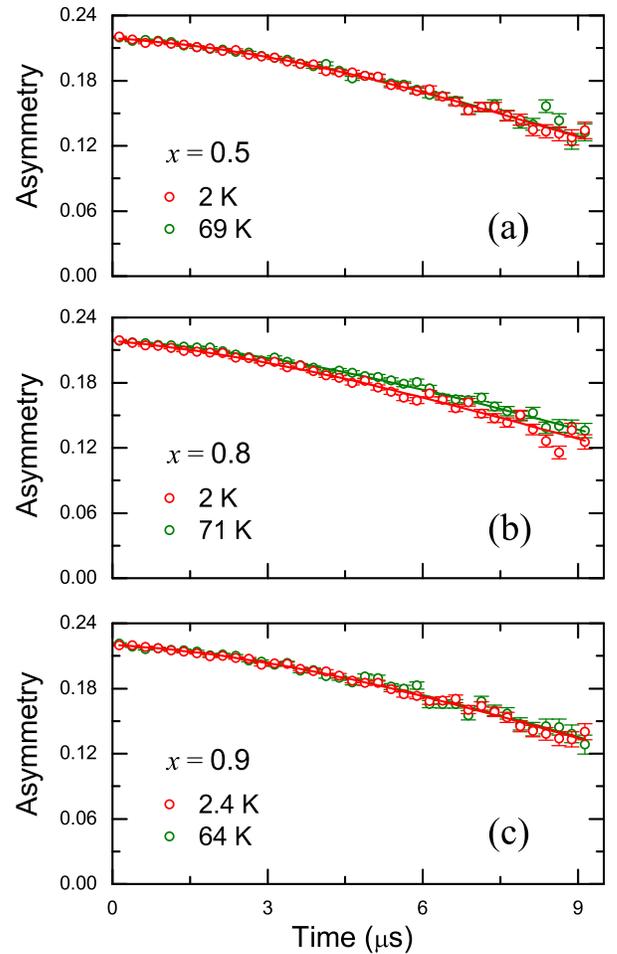}
\caption{(Color online) Representative ZF-$\mu$SR asymmetry spectra of Ba$_{1-x}$K$_{x}$Fe$_2$As$_2$ for
(a) $x \! = \! 0.5$, (b) $x \! = \! 0.8$, and (c) $x \! = \! 0.9$, measured with the sample contained
in a He$^4$ continuous-flow cryostat. The solid curves superimposed on the data points are fits to Eq.~(\ref{eq:Asymmetry}),
which are described in the main text.}
\label{fig2}
\end{figure}   

Single crystal growth of Ba$_{1-x}$K$_{x}$Fe$_2$As$_2$, especially for overdoped compositions, 
is rather challenging due to the vaporization and reaction of K with the alumina crucibles. Our own growth efforts 
failed to obtain uniform crystals with controlled K-contents, and consequently polycrystalline samples were used in 
the present study. The Ba$_{1-x}$K$_{x}$Fe$_2$As$_2$ ($x \! = \! 0.5$, 0.6, 0.7, 0.8, and 0.9) samples were 
synthesized starting with the individual elements. First, purified Fe and As powders, and small pieces of Ba and K were 
mixed and loaded into an alumina crucible inside of a glove box. The alumina crucible was sealed in a Ta 
tube, and the Ta tube subsequently sealed in a quartz tube under 1/3 atmosphere of ultra-high purity argon. 
The ampoule was gradually heated to 750~$^{\circ}$C in a programmable furnace at a ramp rate of 
15~$^{\circ}$C per hour. After staying at 750~$^{\circ}$C for 48 hours, the furnace was shut off. The mixture 
was taken out of the alumina crucible and thoroughly ground to ensure homogeneity, and then pelletized
inside of a glove box. The pellets have a diameter of 1.8 cm and weigh $\sim \! 4.5$~g. 
Each pellet was then loaded into an alumina crucible and sealed in a Ta tube. The final 
sintering was performed at 900~$^{\circ}$C for 48 hours. Room temperature x-ray powder diffraction was 
performed on a PANalytical X'Pert Pro MPD powder x-ray diffractometer using Cu K$\alpha$ radiation. The 
x-ray powder diffraction pattern confirmed that all of the samples are essentially single phase. The 
$T_c$ value of each sample was determined from the temperature dependence of the bulk 
magnetization, measured at an applied magnetic field of 20~Oe with a Quantum Design Magnetic Properties 
Measurement System --- yielding $T_c \! = \! 47$, 30, 20, 11, and 8~K for the 
$x \! = \! 0.5$, 0.6, 0.7, 0.8, and 0.9 samples, respectively. The transition width for all
samples is $\pm 1$~K. These values of $T_c$ agree with previous literature reports.\cite{Johrendt:09,Avci:12}

The ZF-$\mu$SR measurements were performed at TRIUMF in Vancouver, Canada, on the M15 surface positive muon ($\mu^+$) 
beam line, using a Quantum Technology Corp. side-loading, He$^4$ continuous-flow cryostat for measurements
down to $T \! \sim \! 2$~K, and an Oxford Instruments dilution refrigerator for measurements down to
$T \! \sim \! 0.02$~K. In the cryostat the samples were suspended with thin aluminized Mylar tape, and in
the dilution refrigerator mounted on a pure Ag sample holder --- in both cases to avoid a temperature-dependent background 
contribution to the ZF-$\mu$SR signal from materials with electron magnetic dipole moments. 
A ``veto'' detector placed downstream of the sample was used to reject muons that did not stop in 
either the sample or the sample holder. 

An implanted $\mu^+$ precesses
about the local magnetic field $B$ with a Larmor frequency $\omega \! = \! \gamma_\mu B$, where $\gamma_\mu$ is
the muon gyromagnetic ratio. In the SC phase, diamagnetic screening of any external 
magnetic field along the muon spin direction can result in the onset of an
enhanced temperature-dependent relaxation of the ZF-$\mu$SR signal at $T_c$, mimicking the effect of 
induced weak spontaneous internal fields associated with a TRSB pairing state. Even if the mixed TRSB states predicted 
for Ba$_{1-x}$K$_{x}$Fe$_2$As$_2$ occur only at temperatures well below $T_c$, the additional 
temperature-dependent relaxation caused by diamagnetism can mask the contribution of the weak spontaneous fields.
Hence it is crucial to minimize the external field in this type of experiment.
This was achieved by using 3 orthogonal pairs of Helmholtz coils
to compensate for the magnetic field penetrating quartz or pure Si placed at the sample position.
In quartz or Si the positive muon binds to an electron to form the hydrogen-like state muonium (Mu $\equiv \! \mu^+$e$^-$), 
where the muon senses its local environment through the coupled electron. The much larger magnetic moment of the 
electron provides an enhanced sensitivity to internal magnetic fields, such that
the gyromagnetic ratio of Mu is $\gamma_{Mu} \! \sim \! 103 \gamma_{\mu}$. By this method the 
external magnetic field contribution at the sample position was reduced to less than 0.1~Oe.

\begin{figure}
\centering
\includegraphics[width=9cm]{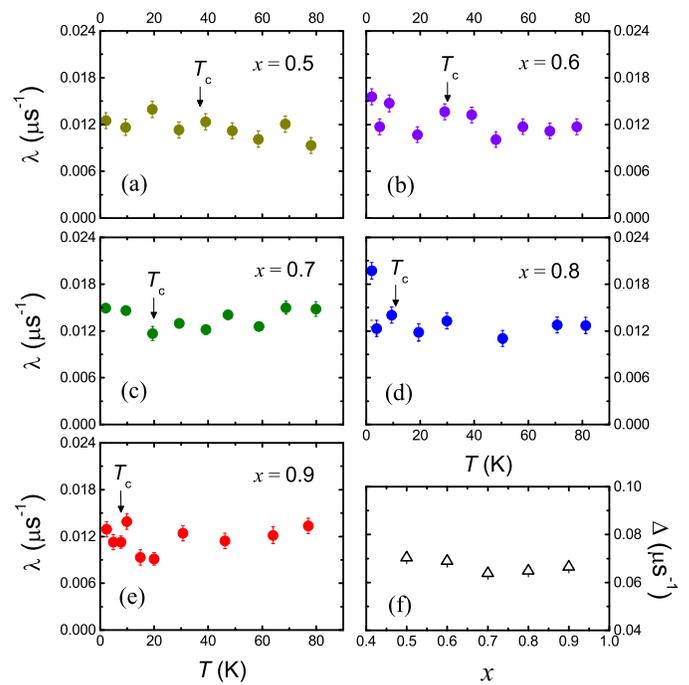}
\caption{(Color online) (a)-(e) Temperature dependence of the exponential relaxation rate $\lambda$, and
(f) the $x$ dependence of $\Delta$, from fits of the ZF-$\mu$SR signals to Eq.~(\ref{eq:Asymmetry}).}
\label{fig3}
\end{figure}

\begin{figure}
\centering
\includegraphics[width=9.5cm]{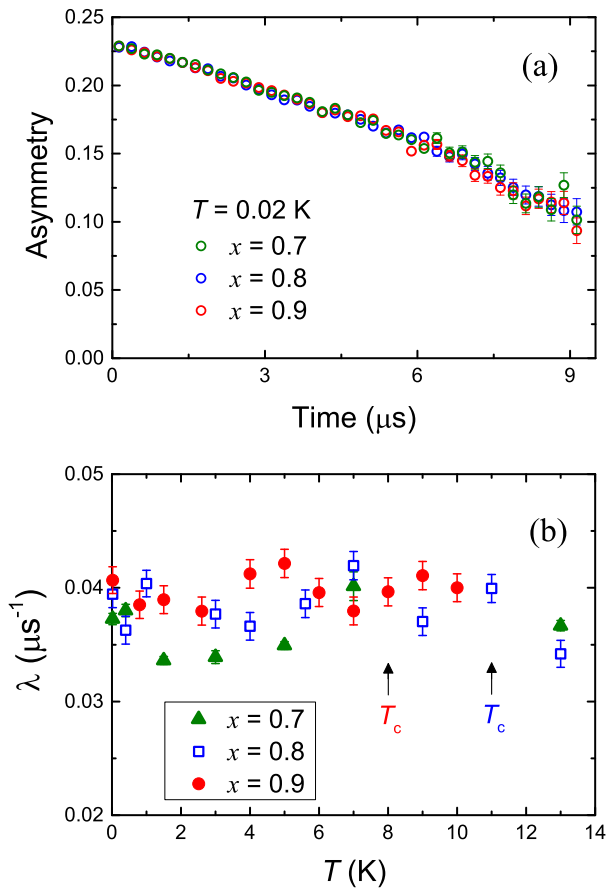}
\caption{(Color online) Results of the low-temperature measurements using a dilution refrigerator. 
(a) ZF-$\mu$SR asymmetry spectra of 
Ba$_{1-x}$K$_{x}$Fe$_2$As$_2$ at $T \! = \! 0.02$~K for $x \! = \! 0.7$, 0.8, and 0.9. (b) Temperature dependence 
of the exponential relaxation rate $\lambda$ from fits of the low-temperature ZF-$\mu$SR signals to Eq.~(\ref{eq:Asymmetry}).}
\label{fig4}
\end{figure}        

Earlier $\mu$SR experiments on single crystals detected a coexistence of 
mesoscopic phase-separated static magnetic order and nonmagnetic/SC regions near 
$x \! = \! 0.5$,\cite{Aczel:08,Goko:09,Park:09} whereas a ZF-$\mu$SR study of 
polycrystalline samples of Ba$_{1-x}$K$_{x}$Fe$_2$As$_2$ reported microscopic coexistence
of magnetism and superconductivity in the underdoped region $x \! \leq \! 0.23$.\cite{Wiesenmayer:11}
Figure~\ref{fig2} shows a comparison of the ZF-$\mu$SR asymmetry spectra of our $x \! = \! 0.5$, 0.8 
and 0.9 polycrystalline samples at a temperature far above $T_c$, and in the superconducting 
phase at $T \! \sim \! 2$~K. The absence of a coherent oscillation in these ZF-$\mu$SR time spectra
is compatible with investigations of the magnetic phase diagram of polycrystalline 
Ba$_{1-x}$K$_{x}$Fe$_2$As$_2$ by other techniques,\cite{Avci:12,Rotter:09} which indicate that 
antiferromagnetic ordering vanishes by $ x \! = \! 0.3$. Furthermore, the lack of an appreciable 
temperature dependence to the ZF-$\mu$SR signal indicates that randomly oriented or isolated 
quasi-static electronic magnetic moments are also absent. 

The solid curves through the data points of the asymmetry spectra in Fig.~\ref{fig2} are fits to
\begin{equation}
A(t) =  A(0) G_{\rm KT}(t) \exp[-\lambda(T) t] \, ,
\label{eq:Asymmetry}
\end{equation}
where $G_{\rm KT}(t) = \left[\frac{1}{3} +\frac{2}{3}(1-\Delta^2 t^2) e^{-\Delta^2 t^2/2}\right]$ is
a ``static Gaussian Kubo-Toyabe function''. It is used here to account for the time 
evolution of the muon-spin polarization caused by the randomly oriented nuclear moments in the sample, 
which generally contribute a temperature-independent Gaussian distribution in field of 
width $\Delta/\gamma_\mu$. For the low-$T$ measurements in the dilution refrigerator there is a small additional
temperature-independent contribution to the ZF-$\mu$SR signal from muons stopping in the sample holder.   
While this component necessarily adds to the sample signal, we find good fits are still achieved using
Eq.~(\ref{eq:Asymmetry}). The fitted value of $\Delta$ is essentially independent of K concentration 
[see Fig.~\ref{fig3}(f)], indicating a minor change in the nuclear dipole contribution. 
The exponential relaxation function in Eq.~(\ref{eq:Asymmetry}) is intended 
to account for any additional sources of internal magnetic field, and unlike $\Delta$ was free to vary 
with temperature in the fits to the ZF-$\mu$SR signals.
As shown in Figs.~\ref{fig3}(a) to \ref{fig3}(e), the exponential relaxation rate $\lambda$ does
not systematically vary with $x$ nor exhibit an abrupt increase at low temperatures indicative 
of spontaneous internal magnetic fields. The small finite value of $\lambda$ may be due to a fluctuating 
muon-nuclear interaction, but is more likely caused by paramagnetic fluctuations of trace amounts of 
an impurity phase --- such as FeAs$_2$, which above $T \! = \! 2$~K is known to cause an exponential 
relaxation of the ZF-$\mu$SR signal.\cite{Baker:08}
 
The larger value of $\lambda$ at $T \! = \! 2$~K for the $x \! = \! 0.8$ sample measured in the
He$^4$ cryostat [see Fig.~\ref{fig3}(d)] is caused by muons stopping upstream of the sample in dense helium gas, 
where the external magnetic field is larger. This is obvious from the low-temperature measurements
carried out using the dilution refrigerator. As shown in Fig.~\ref{fig4}(a), the ZF-$\mu$SR signal at $T \! = \! 0.02$~K does
not vary with $x$. Moreover, there is no onset of an increased relaxation rate of the ZF-$\mu$SR signal
at any temperature below $T_c$.     

Despite the theoretical predictions, the absence of spontaneous internal magnetic fields in our 
measurements suggests that a TRSB mixed symmetry pairing state does not occur at or below $x \! = \! 0.9$.
With this said there are a few possibilities to consider:
(i) The region where the $s \! + \! id$ (or $s \! + \! is$) state is present may be very narrow and fall
between two of the dopings studied here. (ii) There may simply be a first-order phase transition between the pure
symmetry states at all temperatures below $T_c$. (iii) In the coexistence region, the predominant tendency may be states 
with the same phase, such as $s \! + \! d$, which do not break time-reversal symmetry.
Possibilities (ii) and (iii) have been discussed in Ref.~\onlinecite{Fernandes:13} as a 
consequence of the presence of nematic fluctuations. Finally, there is the possibility that formation of
the $s \! + \! id$ state is thwarted by impurity scattering in the real material. In pure 
KFe$_2$As$_2$, substitution of Fe by small concentrations of Co rapidly suppresses $T_c$, 
reminiscent of the high sensitivity of a $d$-wave superconductor to impurity scattering.\cite{Wang:12} 
Likewise, it is possible that the $s \! + \! id$ state is equally sensitive to Ba substitution of K between the 
FeAs layers.    

We thank the staff of TRIUMF's Centre for Molecular and Materials Science for technical assistance,
and C. Wu, R.M. Fernandes, R.F. Kiefl, L. Taillefer, C. Kallin, A. Chubukov, and M. Sigrist  
for informative discussions. The work at TRIUMF was supported by the Canadian Institute of 
Advanced Research, and the Natural Sciences and Engineering Research Council of Canada.
JQY and DGM thank Dr. Chenglin Zhang for his help in synthesis. Work at ORNL was supported 
by the U.S. Department of Energy, Office of Basic Energy Sciences, Materials Sciences and 
Engineering Division.


\begin{thebibliography}{xx}

\bibitem{Hirschfeld:11} P.J. Hirschfeld, M.M. Korshunov, I.I. Mazin, Rep. Prog. Phys. {\bf 74}, 124508 (2011).

\bibitem{Ding:11} H. Ding, K. Nakayama, P. Richard, S. Souma, T. Sato, T. Takahashi, M. Neupane, 
Y.-M. Xu, Z.-H. Pan, A.V. Fedorov {\it et al.}, J. Phys.: Condens. Matter {\bf 23}, 135701 (2011).

\bibitem{Mazin:08} I.I. Mazin, D.J. Singh, M.D. Johannes, and M.H. Du, Phys. Rev. Lett. {\bf 101}, 057003 (2008).

\bibitem{Sato:09} T. Sato, K. Nakayama, Y. Sekiba, P. Richard, Y.-M. Xu, S. Souma, T. Takahashi,
G.F. Chen, J.L. Luo, N.L. Wang, and H. Ding, Phys. Rev. Lett. {\bf 103}, 047002 (2009).

\bibitem{Okazaki:12} K. Okazaki, Y. Ota, Y. Kotani, W. Malaeb, Y. Ishida, T. Shimojima, T. Kiss, S. Watanabe, C.-T. Chen,
K. Kihou {\it et al.}, Science {\bf 337}, 1314 (2012).       

\bibitem{Hashimoto:10} K. Hashimoto, A. Serafin, S. Tonegawa, R. Katsumata, R. Okazaki, T. Saito,
H. Fukazawa, Y. Kohori, K. Kihou, C.H. Lee {\it et al.}, Phys. Rev. B {\bf 82}, 014526 (2010).  

\bibitem{Dong:10} J.K. Dong, S.Y. Zhou, T.Y. Guan, H. Zhang, Y.F. Dai, X. Qiu, X.F. Wang, Y. He,
X.H. Chen, and S.Y. Li, Phys. Rev. Lett. {\bf 104}, 087005 (2010).

\bibitem{Reid:12} J.-Ph. Reid, M.A. Tanatar, A. Juneau-Fecteau, R.T. Gordon, S. Ren\'{e} de Cotret,
N. Doiron-Leyraud, T. Saito, H. Fukazawa, Y. Kohori, K. Kihou {\it et al.}, Phys. Rev. Lett. {\bf 109}, 
087001 (2012).

\bibitem{Wang:12} A.F. Wang, S.Y. Zhou, X.G. Luo, X.C. Hong, Y.J. Yan, J.J. Ying, P. Cheng, G.J. Ye, Z.J. Xiang,
S.Y. Li, X.H. Chen, arXiv:1206.2030.

\bibitem{Abdel:13} M. Abdel-Hafiez, V. Grinenko, S. Aswartham, I. Morozov, M. Roslova, O. Vakaliuk, S. Johnston,
D.V. Efremov, J. van den Brink, H. Rosner {\it et al.}, Phys. Rev. B {\bf 87}, 180507 (2013).

\bibitem{Thomale:11} R. Thomale, C. Platt, W. Hanke, J. Hu, and B.A. Bernevig, 
Phys. Rev. Lett. {\bf 107}, 117001 (2011).

\bibitem{Maiti:11} S. Maiti, M.M. Korshunov, T.A. Maier, P.J. Hirschfeld, and A.V. Chubukov, 
Phys. Rev. Lett. {\bf 107}, 147002 (2011).

\bibitem{Xu:13} N. Xu, P. Richard, X. Shi, A. van Roekeghem, T. Qian, E. Razzoli, E. Rienks, G.-F. Chen,
E. Ieki, K. Nakayama {\it et al.}, arXiv:1308.3888. 

\bibitem{Lee:09} W.-C. Lee, S.-C. Zhang, and C. Wu, Phys. Rev. Lett. {\bf 102}, 217002 (2009). 
 
\bibitem{Stanev:10} V. Stanev and Z. Tesanovi\'{c}, Phys. Rev. B {\bf 81}, 134522 (2010).

\bibitem{Platt:12} C. Platt, R. Thomale, C. Honerkamp, S.-C. Zhang, and W. Hanke,
Phys. Rev. B {\bf 85}, 180502(R) (2012).

\bibitem{Fernandes:13} R.M. Fernandes and A.J. Millis, Phys. Rev. Lett. {\bf 111}, 127001 (2013).

\bibitem{Maiti:13} S. Maiti and A.V. Chubukov, Phys. Rev. B {\bf 87}, 144511 (2013). 

\bibitem{Sigrist:98} M. Sigrist, Prog. Theor. Phys. {\bf 99}, 899 (1998).

\bibitem{Sigrist} M. Sigrist (private communciation).

\bibitem{Heffner:90} R.H. Heffner, J.L. Smith, J.O. Willis, P. Birrer, C. Baines, F.N. Gygax,
B. Hitti, E. Lippelt, H.R. Ott, A. Schenck {\it et al.}, Phys. Rev. Lett. {\bf 65}, 2816 (1990). 

\bibitem{Luke:98} G.M. Luke, Y. Fudamoto, K.M. Kojima, M.I. Larkin, J. Merrin, B. Nachumi,
Y.J. Uemura, Y. Maeno, Z.Q. Mao, Y. Mori {\it et al.}, Nature (London) {\bf 394}, 558 (1998).

\bibitem{Aoki:03} Y. Aoki, A. Tsuchiya, T. Kanayama, S.R. Saha, H. Sugawara, H. Sato, W. Higemoto,
A. Koda, K. Ohishi, K. Nishiyama {\it et al.}, Phys. Rev. Lett. {\bf 91}, 067003 (2003).

\bibitem{Hillier:09} A.D. Hillier, J. Quintanilla, and R. Cywinski, Phys. Rev. Lett. {\bf 102}, 117007 (2009).

\bibitem{Maisuradze:10} A. Maisuradze, W. Schnelle, R. Khasanov, R. Gumeniuk, M. Nicklas, H. Rosner, A. Leithe-Jasper,
Y. Grin, A. Amato, and P. Thalmeier, Phys. Rev. B {\bf 82}, 024524 (2010).

\bibitem{Shu:11} L. Shu, W. Higemoto, Y. Aoki, A.D. Hillier, K. Ohishi, K. Ishida, R. Kadono,
A. Koda, O.O. Bernal, D.E. MacLaughlin, Y. Tunashima {\it et al.}, Phys. Rev. B {\bf 83}, 100504(R) (2011). 

\bibitem{Hillier:12} A.D. Hillier, J. Quintanilla, B. Mazidian, J.F. Annett and R. Cywinski,
Phys. Rev. Lett. {\bf 109}, 097001 (2012).

\bibitem{Biswas:12} P.K. Biswas, H. Luetkens, T. Neupert, T. St\"{u}rzer, C. Baines, G. Pascua, 
A.P. Schnyder, M.H. Fischer, J. Goryo, M.R. Lees {\it et al.}, Phys. Rev. B {\bf 87}, 180503(R) (2013).  

\bibitem{Johrendt:09} D. Johrendt and R. P\"{o}ttgen, Physica~C {\bf 469}, 332 (2009).

\bibitem{Avci:12} S. Avci, O. Chmaissem, D.Y. Chung, S. Rosenkranz, E.A. Goremychkin, J.P. Castellan, 
I.S. Todorov, J.A. Schlueter, H. Claus, A. Daoud-Aladine {\it et al.}, Phys. Rev. B {\bf 85}, 184507 (2012).

\bibitem{Aczel:08} A.A. Aczel, E. Baggio-Saitovitch, S.L. Budko, P.C. Canfield, J.P. Carlo, G.F. Chen,
Pengcheng Dai, T. Goko, W.Z. Hu, G.M. Luke  {\it et al.}, Phys. Rev. B {\bf 78}, 214503 (2008).

\bibitem{Goko:09} T. Goko, A.A. Aczel, E. Baggio-Saitovitch, S.L. Budko, P.C. Canfield, J.P. Carlo, 
G.F. Chen, Pengcheng Dai, A.C. Hamann, W.Z. Hu {\it et al.}, Phys. Rev. B {\bf 80}, 024508 (2009).

\bibitem{Park:09} J.T. Park, D.S. Inosov, Ch. Niedermayer, G.L. Sun, D. Haug, N.B. Christensen,
R. Dinnebier, A.V. Boris, A.J. Drew, L. Schulz {\it et al.}, Phys. Rev. Lett. {\bf 102}, 117006 (2009).

\bibitem{Wiesenmayer:11} E. Wiesenmayer, H. Luetkens, G. Pascua, R. Khasanov, A. Amato, H. Potts,
B. Banusch, H.-H. Klauss and D. Johrendt {\it et al.}, Phys. Rev. Lett. {\bf 107}, 237001 (2011).

\bibitem{Rotter:09} M. Rotter, M. Tegel, I. Schellenberg, F.M. Schappacher, R. P\"{o}ttgen,
J. Deisenhofer, A. G\"{u}nther, F. Schrettle, A. Loidl and D. Johrendt, N. J. Phys. {\bf 11}, 025014 (2009).

\bibitem{Baker:08} P.J. Baker, H.J. Lewtas, S.J. Blundell, T. Lancaster, F.L. Pratt, D.R. Parker, M.J. Pitcher,
and S.J. Clarke, Phys. Rev. B {\bf 78}, 212501 (2008).

\end{thebibliography}
\end{document}